\documentclass[conference]{IEEEtran}
\IEEEoverridecommandlockouts
\usepackage{color}
\usepackage{listings}
\usepackage{amsmath}
\usepackage{amssymb}
\usepackage{amsfonts}
\usepackage{graphicx}
\usepackage{epstopdf}
\usepackage{subfigure}
\usepackage{float}
\usepackage{dsfont}
\usepackage{balance}
\usepackage{array}
\usepackage{steinmetz}
\usepackage{amsthm}
\usepackage{algpseudocode}
\usepackage{algorithm}
\usepackage[mathscr]{euscript}

\newtheorem{theorem}{Theorem}
\begin{document}

\title{ Inference-Based  Resource Allocation \\for  Multi-Cell Backscatter Sensor Networks}
\author{
        Panos N. Alevizos and Aggelos Bletsas
\\
\IEEEauthorblockA{School of Electrical and Computer Engineering (ECE), 
Technical University of Crete, Chania, Greece 73100\\
Email: palevizos$@$isc.tuc.gr, aggelos$@$telecom.tuc.gr } }

\maketitle

\begin{abstract}  
This work studies inference-based resource allocation 
in ultra low-power, large-scale backscatter
sensor networks (BSNs). Several  
   ultra-low cost and power sensor devices (tags) are 
illuminated by a carrier and reflect the measured information towards a wireless
core that uses conventional Marconi radio technology.
The development of multi-cell BSNs requires 
  few  multi-antenna cores  and several low-cost scatter radio devices, targeting
  at maximum possible coverage.
 The average 
signal-to-interference-plus-noise  ratio (SINR)
of   maximum-ratio combining (MRC) and zero-forcing (ZF)  linear detectors is found and harnessed
for frequency sub-channel allocation at tags, exploiting  long-term SINR information. 
The  resource allocation   problem is formulated as an integer programming optimization problem and solved through
the Max-Sum message-passing algorithm. The proposed algorithm is fully parallelizable and adheres to simple
message-passing  update rules, requiring mainly addition and comparison operations. 
In addition, the convergence to the optimal solution is attained within very few iteration steps.
 Judicious simulation 
study  reveals that ZF detector is more suitable for large scale BSNs, capable to 
cancel out the intra-cell interference. It is also found that the proposed  
 algorithm offers at least an order of magnitude decrease  in execution time
 compared to conventional convex optimization methods.
\end{abstract}


\section{Introduction}
\label{sec:intro}

Backscatter sensor networks (BSNs) have   emerged as a promising 
technology for ultra low-cost, large-scale wireless sensor networking
and relevant internet-of-things (IoT) applications \cite{Alevizos_PhD_thesis:17,VaBlLe:08,AlTouBle:18}.
A multi-cell BSN consists of a few interrogators (or cores) that act as fusion centers 
and tags/sensors that are responsible for measuring environmental quantities, 
and transmitting the sensed information towards the cores.
Cores use the conventional  Marconi radio technology with
 front-ends consisting of active filters, mixers, and amplifiers. 
On the other hand, tags utilize scatter radio technology
and rely on the reflection principle:
 Cores emit a continuous sinusoidal wave that illuminates
 the  tags in their vicinity, which in turn use \emph{backscattering}, i.e.,
they modulate information onto the incident signal by alternating a  
radio frequency (RF) transistor switch according to the sensed data.


Current BSNs  are extremely power-limited due to the round-trip nature of backscatter communication;
even for free-space  propagation loss, received power decays with the forth-power
as a function of core-to-tag distance \cite{GrDu:09}. 
Especially for passive tags, the maximum achieved range is on the order of a few meters -- far
 from the required standards of   large-scale, low-cost, ubiquitous sensing applications.
As a result, increasing the coverage, or equivalently,
increasing core-to-tag communication range, is of principle importance in low-cost, low-power BSNs.
To overcome  the limited range issues, prior art in backscatter communications has proposed:   (a) semi-passive tags
and (b) power-limited modulation at tags, such as 
minimum-shift keying (MSK) \cite{VaBlLe:08} or frequency-shift keying  (FSK)  \cite{KiBlSi:13_2}. 
 Especially,  FSK  is ideal for power-limited and
low-bit rate communications and in conjunction  with semi-passive tags,
  offers  promising core-to-tag ranges, on the order of 
hundreds of meters \cite{FaAlBl:15,  AlBlKar:17}. 
In contrast to passive tags,  semi-passive tags are powered through an  external power source, e.g., 
a battery, or a super-capacitor, or other ambient sources (e.g., solar or RF or their combination).

This work examines  a  multi-cell  BSN architecture with joint time-frequency multiple-access. 
A measurement phase is  considered to assess signal-to-interference-plus-noise ratio  (SINR) information
at the tags in the BSN.  The frequency sub-channel allocation at tags changes during the measurement phase and
the goal at each core is to assess  the   average SINR for each pair of neighboring tag-frequency sub-channel.
Subsequently, resource allocation based on a  maximization of a specific metric, involving a function of estimated      average SINR is applied
to find the optimal frequency assignment.

Using the above multi-cell BSN framework, the contributions of this work are summarized as follows:
\begin{itemize}
\item
For zero-forcing (ZF) and maximum ratio combining (MRC) multiuser detection techniques,
 the average received SINR is extracted
and subsequently harnessed to produce
a generic  formulation for resource allocation in BSNs. 
The formulated problem is attacked with a  message-passing   inference algorithm
with simple update rules and convergence to the desired solution within very few
lightweight steps. 
The proposed algorithm is an instance of the Max-Sum algorithm.
\item Judicious simulation study corroborates theoretical findings, showing  that  the proposed 
Max-Sum algorithm offers the same optimal performance with classic convex optimization methods
with reduced computational cost, measured  in terms of execution time in large-scale BSNs. 
\end{itemize}

\emph{Notation}:
The set of real, complex, natural, and binary numbers is denoted 
$\mathds{R}$, $\mathds{C}$, $\mathds{N}$, and $\mathds{B}$, respectively.
Operators   $(\cdot)^{*}$,  $\Re \{ \cdot\}$,  $(\cdot)^{\top}$ ,  $(\cdot)^{\mathsf{H}}$,  
and   $(\cdot)^{\dagger}$
take the conjugate, real part, transpose,  conjugate transpose,  and pseudo-inverse respectively.  
$\mathbf{I}_N$ and $\mathbf{0}_N$  ($\mathbf{1}_N$) represent the $N\times N$ identity matrix and
the   all-zeros  (all-ones) vector of size $N$, respectively.
$ \mathcal{CN}(\boldsymbol{\mu}, \boldsymbol{\Sigma})$ denotes the proper complex Gaussian distribution
with mean $\boldsymbol{\mu}$ and covariance matrix  $\boldsymbol{\Sigma}$.
 $\mathbb{E}[\cdot]$ 
denotes the 
 expectation 
  operator. 

\section{Wireless Scatter Radio System Model}
\label{sec:sys_model}

\begin{figure}[!t]
\centering
   \includegraphics[width=0.95\columnwidth]{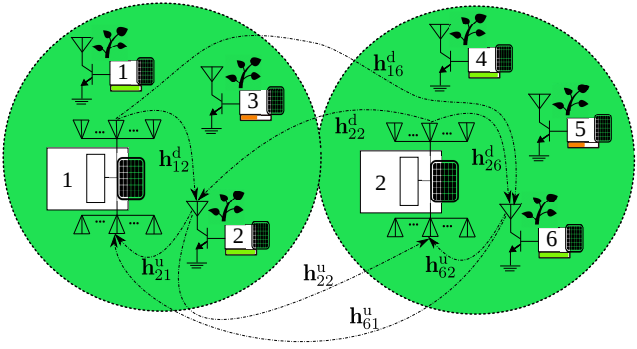}

\caption{A multi-cell BSN  with $B=2$ cores and $K=6$ tags.}
\label{fig:BackscatterSensorNetwork}
\end{figure}

We consider a static multi-cell backscatter sensor network (BSN) consisting of $B$
cores and $K$ sensors/tags; an example is given in Fig.~\ref{fig:BackscatterSensorNetwork}. Each core  is a Marconi radio with separate transmit and receive antennas, e.g.,  a 
software define radio (SDR) reader. 
It is assumed that each core  has $N_{\rm T}$ transmit and $N_{\rm R}$
receive  antennas. 
The set of all cores and tags  is given by    the sets $\mathcal{B}\triangleq \{1,2,\ldots, B\}$ and
 $\mathcal{K} \triangleq \{1,2,\ldots,K\}$, respectively. There exist at total $C$ orthogonal frequency sub-carriers (or sub-channels)
$\left\{ \mathtt{f}^{(1)},\mathtt{f}^{(2)}, \ldots,\mathtt{f}^{(C)} \right\}$, given in an ascending order, indexed by set $\mathcal{C}\triangleq \{1,2,\ldots,C\}$. The distance between core $b$ and tag $k$ is denoted $\mathsf{d}_{kb}$.

Due to  relatively  small symbol rate, $1/T$, delay spread is considered negligible,
and thus, frequency  non-selective (flat) fading channel \cite{Goldsmith:05} is assumed across   
core-to-tag links $(b,k)$ and tag-to-core links $(k,b)$, $b \in \mathcal{B}$, $k \in \mathcal{K}$. 
For outdoor environments,  it is customary to assume  strong  line-of-site components, and thus,
 the  {baseband} complex channel  responses  for downlink $(b,k)$ and  uplink $(k,b)$ are, respectively,
  given by: 
  \begin{align}
 \mathbf{h}_{bk}^{\rm d}      \sim \mathcal{CN}\left(\sqrt{\frac{\kappa_{bk}^{\rm d}}{\kappa_{bk}^{\rm d} + 1}}  \sigma_{bk} \mathbf{e}_{bk}^{\rm d} ,
\frac{\sigma_{bk}^2}{\kappa_{bk}^{\rm d} + 1} \mathbf{I}_{N_{\rm T}}\right) ,
 \label{eq:small_scale_fading_bk}  \\
 \mathbf{h}_{kb}^{\rm u} \sim \mathcal{CN}\left(\sqrt{\frac{\kappa_{kb}^{\rm u}}{\kappa_{kb}^{\rm u}+ 1}}  \sigma_{kb}  \mathbf{e}_{kb}^{\rm u}  ,
\frac{\sigma_{kb}^2}{\kappa_{kb}^{\rm u} + 1}  \mathbf{I}_{N_{\rm R}} \right), \label{eq:small_scale_fading_kb}
\end{align}
where $\kappa_{bk}^{\rm d}$ and $\kappa_{kb}^{\rm u}$ denote the   ratio between the power in the direct  path and the power in the  scattered paths of  core-to-tag   link $(b,k)$ and  tag-to-core  link $(k,b)$, respectively.   $\mathbf{e}_{bk}^{\rm d}$ and
$\mathbf{e}_{kb}^{\rm u}$   are the antenna steering vectors, 
depending, respectively, on the angle-of-departure (AoD) and angle-of-arrival (AoA) between core $b$ and tag $k$.
$\frac{1}{N_{\rm T } }\mathbb{E}[ \|\mathbf{h}_{bk}^{\rm d} \|_2^2 ] =  \sigma_{bk}^2 $
and $\frac{1}{N_{\rm R } } \mathbb{E}[ \|\mathbf{h}_{kb}^{\rm u} \|_2^2 ] =     \sigma_{kb}^2 $
denote the normalized channel powers of   core-to-tag   link    $(b,k)$  and  tag-to-core  link $(k,b)$, respectively. 
Both downlink  $ \{\mathbf{h}_{bk}^{\rm d} \}_{(b,k) \in \mathcal{B} \times \mathcal{K}} $ and 
 uplink $\{ \mathbf{h}_{kb}^{\rm u}
 \}_{(b,k) \in \mathcal{B} \times \mathcal{K}}$
 channel gains are assumed uncorrelated of each other and across any  pair $(b,k)$,  
changing independently every $T_{\rm coh}$
seconds, where $T_{\rm coh}$ is the channel coherence time.
%
The normalized channel powers
 depend on large-scale path-gain (inverse of path-loss) which
is assumed the same for links $(b,k)$ and $(k,b)$, i.e., $\sigma_{bk}= \sigma_{kb}$.

 When a tag $k \in \mathcal{K}$ reflects information, it backscatters a packet of $M$ symbols, where each symbol 
 within a packet takes   values  $\{\pm1\}$. $M_{\rm tr}$ symbols  form a preamble, dedicated for training and
the rest $M_{\rm d} = M - M_{\rm tr} $ are  data symbols.
The set of available training sequences $\mathcal{X}_{\rm tr} =  \left\{\mathbf{x}^{(1)}, 
\mathbf{x}^{(2)}, \ldots, \mathbf{x}^{(M_{\rm tr})} \right\} \subset \{\pm1\}^{M_{\rm tr}}$, i.e.,
 set $\mathcal{X}_{\rm tr}  $ contains $ M_{\rm tr}$ sequences,
each of dimension $ M_{\rm tr}$.
The sequences in set  $\mathcal{X}_{\rm tr}  $ are orthogonal, i.e., 
 $\left(\mathbf{x}^{(m)}\right)^{\top}  \mathbf{x}^{(m')}  = 0$, for $m \neq m'$.
The training sequence  assigned to a tag $k$ has been decided a-priori and  is assumed fixed.
 In a practical scenario, to reduce the training interference, the same training sequence has to be reused across tags that are far
apart.   
Let $\mathcal{M}_{\rm tr}\triangleq \{1,2,\ldots, M_{\rm tr}\}$ and
$\mathcal{M}_{\rm d} \triangleq  \{M_{\rm tr}+1 , M_{\rm tr}+2, \ldots, M\}$.
The set of tags assigned to the $m$-th training sequence is expressed as
$\mathcal{K}_{\mathcal{M}_{\rm tr}}(m) \! \triangleq \! \{k \in \mathcal{K} \!: \text{tag } k \text{ uses training sequence } \mathbf{x}^{(m)}  \} ,$  the set of tags assigned to $c$-th frequency sub-channel is
$\mathcal{K}_{\mathcal{C}}(c) \!  \triangleq \! \{k \in \mathcal{K}: \text{tag } k \text{ assigned to  sub-carrier } \mathtt{f}^{(c)}   \} $,   
and the tags belonging to cell $b$ are given by the following set
$\mathcal{K}_{\mathcal{B}}(b)  \! \triangleq \! \{k \in \mathcal{K} : b =  {\rm arg} \min_{b' \in \mathcal{B}} \mathsf{d}_{b'k} \}$.
It is not difficult to see that the set $\mathcal{K}$
can be partitioned as $\mathcal{K} = \bigcup_{b \in \mathcal{B} } \mathcal{K}_{\mathcal{B}}(b)$,
where $\mathcal{K}_{\mathcal{B}}(b) = \bigcup_{m \in \mathcal{M}_{\rm tr}} \mathcal{K}_{bm}$,
with  $\mathcal{K}_{bm} \triangleq  \mathcal{K}_{\mathcal{B}}(b)  \cap \mathcal{K}_{\mathcal{M}_{\rm tr}}(m) $
and each $ \mathcal{K}_{bm} $ can be further partitioned as
  $ \mathcal{K}_{bm}  =  \bigcup_{c \in \mathcal{C}} \mathcal{K}_{bmc}$,
where  $ \mathcal{K}_{bmc} =  \mathcal{K}_{bm} \cap \mathcal{K}_{\mathcal{C}}(c)$. 
Easily follows that  $ \mathcal{K}_{bmc}$ and $\mathcal{K}_{b'm'c'}$ are disjoint 
if $b\neq b'$ or
$m\neq m'$ or $c\neq c'$.

To minimize intra-cell interference,  
    the
tags sharing the same training sequence are 
configured to backscatter with different switching frequency rate (i.e., sub-carrier). Namely,
tags within the same cell are assigned to a  unique pair $ (m,c) \in \mathcal{M}_{\rm tr}\times  \mathcal{C} $ 
(i.e.,  $|\mathcal{K}_{bmc}| \leq 1$).



For any core  pair $(b,b') \in \mathcal{B}\times \mathcal{B}$, $b'$ emits  a continuous sinusoidal wave with baseband representation 
$ \mathsf{q}_{bb'}(t) = \sqrt{ \frac{ P_{b'}}{N_{\rm T}}} \, \mathsf{e}^{-\mathsf{j}(2 \pi \Delta F_{bb'} t+ \phi_{bb'})}$,
where $P_{b'}$ is the total transmission power of core $b'$, $\Delta F_{bb'}$ 
 and $\phi_{bb'}$ are the  carrier frequency  offset (CFO) and carrier  phase offset (CPO) between core's $b'$ transmit circuity  and core's $b$ receive circuitry, which both are  deterministic and varying very slowly over time. 
 For a tag-core pair $(k,b) \in \mathcal{K} \times \mathcal{B}$,
the superposition of emitted signals from cores $b' \in \mathcal{B}$  propagated across 
downlink channels $\{\mathbf{ h}_{b'k}^{\rm d} \}_{b'\in \mathcal{B}}$
 impinges on the antenna of tag $k$, i.e., tag  $k$ receives 
 $
\sum_{b'=1}^{B} \mathsf{q}_{bb'}(t)   \mathbf{1}_{N_T}^{\top}   \mathbf{h}_{b'k}^{\rm d}.
$
According to tag's $k$ measured data, 
an alternation of its antenna load is performed   producing the following
backscattered signal towards core $b$ \cite{KiBlSi:13_2, FaAlBl:15}:
\begin{align}
 \mathsf{u}_{kb}(t) = & \,{\rm DC}_{k} +  \sum_{b'=1}^{B}   \sqrt{ \frac{ P_{b'}}{N_{\rm T}}}  \,\mathsf{e}^{-\mathsf{j}(2 \pi \Delta F_{bb'}t+ \phi_{bb'})}\mathbf{1}_{N_{\rm T}}^{\top}  \mathbf{h}_{b'k}^{\rm d}  \eta_k      \cdot \nonumber \\
  \cdot &  ~ \frac{(\Gamma_{k,0}-\Gamma_{k,1})}{2}\sum_{i=0}^{M-1} x_{k,i+1}\, \mathsf{v}_{kb}(t - iT),
\label{eq:backscatter_signal_tag_k}
\end{align}
where $ {\rm DC}_{k}$ is a DC term, independent on time  $t$,
depending solely on antenna structural mode and the loads of tag $k$ \cite{BleDiSah:10}.
  ${\eta}_k $ is the scattering efficiency, 
remaining constant within packet duration,
$\Gamma_{k,0}$ and $\Gamma_{k,1}$ are the (load-dependent) reflection coefficients,
and $\mathsf{v}_{kb}(t)$  is the reflected waveform  of tag $k$. 
Incorporating the alternation of tag's $k$  switch, waveform
 $\mathsf{v}_{kb}(t)$ is the fundamental frequency component 
of a $50 \%$ duty cycle square waveform of period $1/ f_k$ and bit (or symbol) duration $T$, i.e.,
$
 \mathsf{v}_{kb}(t)=\frac{4}{\pi}\mathsf{cos}\!\left(2\pi f_k t  + \Phi_{kb}\right) \Pi_{T}(t),
\label{eq:tag_k_waveform}
$
 where  $\Pi_{T}(t)$ is $1$ when $t \in [0,T)$, and $0$ otherwise.
$f_k$ and $\Phi_{kb}$  are the generated frequency of tag $k$  and the random
   phase mismatch  between tag $k$  and core $b$, respectively.  The latter depends on the 
   channel propagation delay and the transmitted signal and is modeled as  uniform  RV in $[0,2\pi)$.
If the $m$-th training sequence and the $c$-th frequency sub-channel are assigned to tag $k$,
then $ \{x_{k,i}\}_{i=1}^{M_{\rm tr}} = \mathbf{x}^{(m)} $  
  and $f_k= \mathtt{f}^{(c)}$ holds.
%

%
Core $b$ receives the superposition 
of  $\{ \mathsf{u}_{kb}(t)\}_{k \in\mathcal{K}}$,  propagated  by uplink channels $\left\{ \mathbf{h}^{\rm u}_{kb}\right\}_{k \in \mathcal{K}} $, i.e.,%
\begin{equation}
\widetilde{\mathsf{\mathbf{y}}}_b(t) = \sum_{k \in \mathcal{K}}  \mathbf{h}^{\rm u}_{kb} \mathsf{u}_{kb}(t) + \mathbf{w}_b(t),
\label{eq:Rx_signal_core_b}
\end{equation}
where each component  of $\mathsf{\mathbf{w}}_b(t)$ is independent circularly symmetric, complex
      Gaussian random process with flat power
spectral density  $N_0$ over $ [- \mathtt{B}_b,  \mathtt{B}_b]$  band, and zero otherwise, 
with
parameter $\mathtt{B}_b$ denoting the receiver bandwidth at core $b$. 

Signal in~\eqref{eq:Rx_signal_core_b} contains  CFO, CPO, and DC terms. Assuming
perfect compensation for the  slow-varying  $\Delta F_{bb'}$ and $\phi_{bb'}$, 
the DC term is compensated by removing the time-average of $\widetilde{\mathbf{y}}_b(t)$
from the signal itself, i.e.,
   $\mathbf{y}_b(t) = \widetilde{\mathbf{y}}_b(t) - \int_{\mathcal{T}_0} \widetilde{\mathbf{y}}_b(t) \mathsf{d}t$,
 where $\mathcal{T}_0 \supseteq [0, MT]$ is the time processing interval.
  Hence, 
abbreviating $\forall b \in \mathcal{B}$, $\forall k \in \mathcal{K}$,
\begin{equation}
\mathbf{g}_{kb}  \triangleq  \frac{ \sum \limits_{b'=1}^{B}   \! 
 \sqrt{ \frac{P_{b'}}{ N_{\rm T}}}
\mathbf{1}_{N_{\rm T}}^{\top}
  \mathbf{h}_{b'k}^{\rm d} \eta_k  \, (\Gamma_{k,0}-\Gamma_{k,1})\, \mathbf{h}^{\rm u}_{kb}  }{
   \frac{\pi}{2}} 
\label{eq:g_kb}
\end{equation}
and plugging Eqs.~\eqref{eq:backscatter_signal_tag_k} and~\eqref{eq:g_kb}
in~\eqref{eq:Rx_signal_core_b} ,
 the DC-blocked, CFO-free signal   $  \mathbf{y}_b(t) $ can be written as:
\begin{align}
\mathbf{y}_b(t) = &   \sum_{k \in \mathcal{K} }  \mathbf{g}_{kb}    \sum_{i=0}^{M-1} x_{k,i+1}\, 
\mathsf{cos}\! \left[2 \pi f_k (t - iT )
   + \Phi_{kb}\right]  +   \mathbf{w}_b(t) ,
\label{eq:DC_blocked_Rx_signal_core_b}
\end{align}
for $ t \in \mathcal{T}_0$.
The received signal $\mathbf{y}_b(t)$ passes through correlators and the outcome is sampled
at time instants $\{iT~ |~ i=1,2,\ldots, M\}$. 
 \begin{theorem} \normalfont \label{theorem:multiuser_baseband_equivalent}
If (a) the frequency sub-carriers  satisfy $\mathtt{f}^{(c)}=  \frac{ l_c}{T} , ~~ \forall c\in \mathcal{C},
$ for some $l_c \in \mathds{N}$ and (b) SDR bandwidth satisfies $\mathtt{B}_b \gg \frac{1}{T}$, then  the 
discrete baseband  equivalent signal vector at core $b$   
can be written as:
\begin{equation}
\mathbf{r}_{b,i}^{(c)}=  \sum_{   k \in \mathcal{K}_{\mathcal{C}}(c)   } \boldsymbol{ \xi}_{kb}^{(c)} \,  {x}_{k,i} + 
     \mathbf{n}_{b}^{(c)}, ~ i =1,2,\ldots, M,
\label{eq:multi_user_baseband}
\end{equation}
where  $\boldsymbol{\xi}_{kb}^{(c)} \triangleq   \mathbf{g}_{kb} \,  \sqrt{\frac{ T}{2}}\, \mathsf{cos}(\Phi_{kb})  $
if  $ k \in \mathcal{K}_{\mathcal{C}}(c) $ and zero  otherwise. Vector   $\boldsymbol{\xi}_{kb}^{(c)}$ is
 the compound (uplink) channel between tag $k$ and core $b$ at the output of $c$-th frequency matched filter,
 incorporating microwave and wireless propagation parameters,
while  noise vector  $  \mathbf{n}_{b}^{(c)} \sim \mathcal{CN}(\mathbf{0}_{N_{\rm R}},  \sigma_b^2 \, \mathbf{I}_{N_{\rm R}}) $,
with  $\sigma_b^2=N_0$.
\end{theorem}

\section{Linear Detection and SINR Calculation}
\label{sec:multitag detection and SINR}

Assuming knowledge of  $\{\boldsymbol{ \xi}_{kb}^{(c)}\}_{c \in \mathcal{C}}$
for all $k \in \mathcal{K}_{\mathcal{B}} (b)$,
a multi-tag linear detector is applied as a linear operator $\mathbf{a}_{kb}^{(c)}$
on the received vector $ \mathbf{r}_{b,i}^{(c)}$ , i.e., 
\begin{align}
z_{k,i} ~&= (\mathbf{a}_{kb}^{(c)})^{\mathsf{H}} \mathbf{r}_{b,i}^{(c)} \overset{(a)}{=} 
  (\mathbf{a}_{kb}^{(c)})^{\mathsf{H}}   \Big(  {\boldsymbol {\xi}}_{kb}^{(c)} x_{ki} 
+ \sum_{  \begin{smallmatrix} m' \in \mathcal{M}_{\rm tr} \\  k' \in \mathcal{K}_{bm'c} \backslash k \end{smallmatrix}  }
 {\boldsymbol {\xi}}_{k'b}^{(c)}  x_{k'i} 
\nonumber \\
&  + ~
\sum_{b' \neq b } \sum_{  \begin{smallmatrix}  m' \in \mathcal{M}_{\rm tr} \\  k' \in \mathcal{K}_{b'm'c} \end{smallmatrix}   }   \boldsymbol {\xi}_{k'b}^{(c)}  x_{k'i}  +  \mathbf{n}_{b}^{(c)}  
 \Big),
\label{eq:detection_baseband_decomposed}
\end{align}
where in $(a)$,  the set $\mathcal{K}_{\mathcal{C}}(c)$  is partitioned
to following disjoint sets: $(i)$ the desired user $k$, $(ii)$ the intra-cell interferers $\mathcal{I}_{\rm in}(k) \triangleq\bigcup_{m' \in \mathcal{M}_{\rm tr}}  \mathcal{K}_{bm'c} \backslash k$ (i.e. the tags  in cell $b$  assigned to the  $c$-th
frequency  sub-channel, excluding $k$)
and $(iii)$ the inter-cell interferers $\mathcal{I}_{\rm out}(k) \triangleq \bigcup_{b' \neq b} \bigcup_{m' \in \mathcal{M}_{\rm tr}}  \mathcal{K}_{b'm'c}$ (i.e. the tags from all cells $b'$ except cell $b$ assigned to the  $c$-th
frequency  sub-channel).
The estimate of $x_{k,i}$ is given by 
$ \widehat{x}_{k,i} = \mathsf{sign}(\Re \{z_{k,i}\}),~ i \in \mathcal{M}_{\rm d}$ \cite{Goldsmith:05}.

Two linear detection techniques are examined 
 for symbol $x_{k,i}$, $i \in \mathcal{M}_{\rm d} $:  maximum-ratio  combining (MRC)
and   zero-forcing (ZF). 
 For   MRC detection, vector $\mathbf{a}_{kb}^{(c)} =    {\boldsymbol{\xi}}_{kb}^{(c)}$. On the other hand,  
for ZF detection, core $b$ partitions the tags in the cell according to their utilized frequency sub-channels;  for    sub-channel $\mathtt{f}^{(c)}$
the following matrix is formed: 
$\mathbf{P}^{(c)}_b = \left[ {\boldsymbol{\xi}}_{l_1 b}^{(c)}~ {\boldsymbol{\xi}}_{l_2 b}^{(c)}~\ldots
~ {\boldsymbol{\xi}}_{ l_{K_{bc}}b}^{(c)}\right]$,
where $ \mathcal{K}_{\mathcal{B}}(b) \cap  \mathcal{K}_{\mathcal{C}}(c)  = \{l_1, l_2, \ldots l_{K_{bc}}\}  $ and
  $K_{bc} = |  \mathcal{K}_{\mathcal{B}}(b) \cap  \mathcal{K}_{\mathcal{C}}(c)|$.
The final  ZF   operator is given by  $(\mathbf{a}_{kb}^{(c)})^{\mathsf{H}} = \left[ (\mathbf{P}^{(c)}_b)^{\dagger}\right]_{q,:}$, where   the $ q$-th element of set  $ \mathcal{K}_{\mathcal{B}}(b) \cap  \mathcal{K}_{\mathcal{C}}(c)$  satisfies  $l_q = k$.
Note that ZF detector tries to mitigate the intra-cell interference coming from the tags in cell $b$
  using the same frequency sub-channel with tag $k$. It holds that ZF can fully mitigate  intra-cell interference, provided that $N_{\rm R} \geq K_{bc} $,  i.e., $(\mathbf{a}_{kb}^{(c)})^{\mathsf{H}}  {\boldsymbol {\xi}}_{k'b}^{(c)} =0 $
for $k' \in ( \mathcal{K}_{\mathcal{B}}(b) \cap  \mathcal{K}_{\mathcal{C}}(c)) \backslash{k}$.

Core $b$ treats the channel vectors within its cell (i.e., $ \{\boldsymbol{\xi}_{kb}^{(c)} : 
\{k\} \cup \mathcal{I}_{\rm in}(k) \}$) as known
and the  terms  of inter-cell interference  and noise  in~\eqref{eq:detection_baseband_decomposed} 
are  considered as random.
Thus, using the independence of  zero-mean  $\{ \boldsymbol{\xi}_{k'b}^{(c)} \}_{ k' \in
\mathcal{I}_{\rm out}(k)}$ \cite{Alevizos_PhD_thesis:17},  $\{ x_{k,i}\}_{k \in \mathcal{K}}$, 
and $ \mathbf{n}_{b}^{(c)}$, the instantaneous received signal-to-interference-plus-noise (SINR)  for tag $k \in \mathcal{K}_{bmc}$   is given  by:
\begin{equation}
{\rm SINR}_{kb}^{(c)} =  
\frac{  \left| (\mathbf{a}_{kb}^{(c)})^{\mathsf{H}}    {\boldsymbol {\xi}}_{kb}^{(c)}  \right|^2  }{
 I_{kb}^{1(c)} +  I_{kb}^{2(c)}   +   \sigma_b^2\,\| \mathbf{a}_{kb}^{(c)}\|_2^2
 },
\label{eq:SINR_kb_c}
\end{equation}
where intra-cell ($I_{k,b}^{1(c)}$) and inter-cell ($I_{k,b}^{2(c)}$) interference terms 
are given by:
$
 I_{kb}^{1(c)}  =  \!\! \sum_{k' \in \mathcal{I}_{\rm in}(k)}
\!  \left|(\mathbf{a}_{kb}^{(c)})^{\mathsf{H}}  {\boldsymbol {\xi}}_{k'b}^{(c)}  \right|^2$ and
$ I_{kb}^{2(c)}  =  \! \sum_{k' \in \mathcal{I}_{\rm out}(k) }\! 
(\mathbf{a}_{kb}^{(c)})^{\mathsf{H}}   \mathbf{C}_{\boldsymbol {\xi}_{k'b}^{(c)} }   \mathbf{a}_{kb}^{(c)},
\label{eq:SINR_intercell_interference}
$
respectively.  Matrix $  \mathbf{C}_{\boldsymbol {\xi}_{k'b}^{(c)} }  $ is the
 covariance of $\boldsymbol {\xi}_{k'b}^{(c)}$, given in  \cite[Eq.~(5.20)]{Alevizos_PhD_thesis:17}.


The SINR in~\eqref{eq:SINR_kb_c} depends on $\{ {\boldsymbol{\xi}}_{k'b}^{(c)} : k' \in \{k\} \cup
 \mathcal{I}_{\rm in}(k) \cup  \mathcal{I}_{\rm out}(k) \}$
which in turn changes every coherence period. 
To apply robust frequency  allocation,   an average SINR  calculation has to be conducted across many
wireless channel  and frequency allocation realizations.
To this end, a measurement procedure is employed by all cores and
tags to obtain long-term SINR information for subsequent frequency allocation.

Each tag is assigned to a fixed preamble sequence across the whole measurement
phase, while the frequency channel allocation of each tag changes in a per frame basis.
Let us denote $J$ the total number of  frames for the measurement phase, with set $\mathcal{J}\triangleq \{1,2,\ldots J\}$.
 For  a tag $k \in \mathcal{K}_{\mathcal{B}}(b)$,
 let us denote $\mathcal{J}_{kb}^{(c)}$ the measurement indices  for  which tag $k$ 
is assigned to sub-channel  $\mathtt{f}^{(c)}$. For each tag $k \in \mathcal{K}_{\mathcal{B}}(b)$ 
core $b$ calculates the received  SINR in~\eqref{eq:SINR_kb_c} for the frames indexed by the set  $\mathcal{J}_{kb}^{(c)}$,  denoted as ${\rm SINR}_{kb}^{(c)}[j]$,
forming the set    $\left\{{\rm SINR}_{kb}^{(c)}[j] :j \in  \mathcal{J}_{kb}^{(c)}\right \} $.

At the end, for each tag  $k \in \mathcal{K}_{\mathcal{B}}(b)$,
an estimate of average SINR at core $b$ for the $c$-th sub-channel  is obtained as:
\begin{equation} 
\overline{\mathtt{SINR}}_{kb}^{(c)}  = \frac{1}{\left| \mathcal{J}_{kb}^{(c)} \right|} \sum_{j \in \mathcal{J}_{kb}^{(c)} }  {\rm SINR}_{kb}^{(c)}[j].
\label{eq:avg_SINR_kb_c}
\end{equation}

\section{Frequency Allocation Based On Max-Sum Message-Passing}
\label{sec:res_aloc}

%

After obtaining the average SINR estimates $\overline{\mathtt{SINR}}_{kb}^{(c)}$
for all tuples $    \{ (k,b,c) \in \mathcal{K}  \times  \mathcal{B} \times  \mathcal{C} : k\in \mathcal{K}_{\mathcal{B}}(b)\} $,
all cores try to obtain a frequency sub-channel--tag assignment that maximizes a  specific metric involving
the estimated average received SINRs.   
Using the average received SINR instead of
instantaneous SINR, the impact of  random intra- and inter- interference is averaged out, and thus, the optimization 
problem can be decoupled to $B$ parallel sub-problems across all cores.

The proposed formulation to obtain the optimal tag--frequency sub-channel
association is expressed at each core individually  through the following optimization problem:
\begin{subequations}\label{eq:optimization_core_b}
\begin{align}
~~~ \underset{}{\rm maximize  }  &~ ~
  \sum_{k \in   \mathcal{K}_{\mathcal{B} }(b) } \sum_{c \in   \mathcal{C}} \mathsf{g}\!\left({\overline{\mathtt{SINR}}}_{kb}^{(c)} \right) \cdot v_{kc} \label{eq:optimization_core_b_1} \\
~~~ {\rm subject~to} & ~~ \sum_{k \in \mathcal{K}_{bm}} \!\! v_{kc} \leq  1, ~\forall (m,c)\in \mathcal{M}_{\rm tr } \times \mathcal{C}  ,   \label{eq:optimization_core_b_2} \\
~~&~~~  \sum_{c \in \mathcal{C}}  v_{kc} =  1, ~\forall k \in      \mathcal{K}_{\mathcal{B}} (b),    \label{eq:optimization_core_b_4} \\
~~~& ~~~~    \,  {v_{kc} \in \mathds{B}, ~\forall (k,c) \in    \mathcal{K}_{\mathcal{B}} (b)  \times \mathcal{C}}, \label{eq:optimization_core_b_5}
\end{align}
\end{subequations} 
 where $\mathsf{g} :\mathds{R}^{+} \longrightarrow \mathds{R}$ is an arbitrary increasing function. 
Resource allocation variables $ v_{kc}$ indicate  whether tag  $k \in \mathcal{K}$ backscatters on the 
$c$-th  frequency sub-channel  ($ v_{kc}=1$)   or not  ($ v_{kc}=0$).
Constraint~\eqref{eq:optimization_core_b_2} imposes that a frequency sub-channel $\mathtt{f}^{(c)}$ can be assigned to at most one tag in $\mathcal{K}_{bm}$,
$ \forall m \in \mathcal{M}_{\rm tr}$, 
Constraint~\eqref{eq:optimization_core_b_4} dictates that
  each tag has to be assigned to   one frequency sub-channel.
From a practical point of view,    constraint~\eqref{eq:optimization_core_b_2} 
  offers  intra-cell pilot  interference cancellation 
by assigning each  tag in cell $b$ to a unique pair $(m,c) \in \mathcal{M}_{\rm tr} \times \mathcal{C}$ of sequence
$\mathbf{x}^{(m)}$  and frequency sub-channel $\mathtt{f}^{(c)}$, causing orthogonal training transmissions. 
In doing so,
the channel estimate obtained for the tag $k \in \mathcal{K}_{bmc}$ is contaminated only by the 
tags from other cells that use the same frequency sub-channel $\mathtt{f}^{(c)}$ (i.e., $k\in  \mathcal{I}_{\rm out}(k)$). 
 For   set  $\mathcal{K}_{\mathcal{B} }(b) $, the corresponding  assignment matrix  is defined as
 $\mathbf{V}_{b} \triangleq \{v_{kc}:   \forall (k,c) \in  \mathcal{K}_{\mathcal{B} }(b)   \times \mathcal{C} \}$.
The following functions are also defined:
\begin{align}
\mathsf{G}(\mathbf{V}_{b}) 
\triangleq & ~ \sum_{k \in     \mathcal{K}_{\mathcal{B}}(b)  }  \sum_{c \in   \mathcal{C}} \mathsf{G}_{kc}(v_{kc}) \\
\mathsf{G}_{kc}(v_{kc}) \triangleq & ~ \mathsf{g}\Big(\overline{\mathtt{SINR}}_{kb}^{(c)} \Big)  \cdot  v_{kc}  \label{eq:G_kc_factors} \\
 \mathsf{p}_{mc} \!\left(\{ v_{kc}\}_{k \in \mathcal{K}_{bm}} \right) \triangleq & ~
\mathbb{I}\! \left\{    \sum_{k \in \mathcal{K}_{bm}}  v_{kc} \leq1 \right\}, ~ (m,c) \in  \mathcal{M}_{\rm tr} \times \mathcal{C} \\
\mathsf{h}_k \!\left(\{ v_{kc}\}_{c \in \mathcal{C}} \right)
 \triangleq & ~ \mathbb{I}\! \left\{     \sum_{c \in \mathcal{C}}  v_{kc} = 1 \right\},  ~  k \in \mathcal{K}_{\mathcal{B}}(b)
\end{align}
where the last two functions (factors) are
associated with  constraints~\eqref{eq:optimization_core_b_2} 
and~\eqref{eq:optimization_core_b_4}, respectively;   
for a statement $X$, function  $\mathbb{I}\{X\} $ is the max-indicator function defined as
$
\mathbb{I}\{X\}  =0$, if $X$ is true, and $\mathbb{I}\{X\}  =-\infty$
if $X$ is false.

The   integer programming problem in~\eqref{eq:optimization_core_b} belongs to the class of 
maximum weighted matching problems \cite{ BaBoChZe:08} that can be  solved through the Max-Sum
algorithm.  
 It can be shown that the   problem in Eq.~\eqref{eq:optimization_core_b}
is equivalently expressed as:
\begin{align}
  \max_{\mathbf{V}_{b} \in \mathds{B}^{|\mathcal{K}_{\mathcal{B}}(b)| \times | \mathcal{C}|}}  & \! \left\{ 
\mathsf{G}(\mathbf{V}_{b})   + \sum_{ \begin{smallmatrix} m \in \mathcal{M}_{\rm tr} \\                                                            
c\in \mathcal{C} \end{smallmatrix} }  \mathsf{p}_{mc} \!\left(\{ v_{kc}\}_{k \in \mathcal{K}_{bm}} \right)   \right.\nonumber \\
+&  \left.
\sum_{k \in \mathcal{K}_{\mathcal{B}}(b) }\mathsf{h}_k \!\left(\{ v_{kc}\}_{c \in \mathcal{C}} \right)  \right\}.
\label{eq:optimization_core_b_reformulation}
\end{align}
The above  (unconstrained) maximization problem is equivalent to the constrained problem~\eqref{eq:optimization_core_b}
because the constraints in~\eqref{eq:optimization_core_b_2} and~\eqref{eq:optimization_core_b_4}
are imposed   through indicator functions $\{\mathsf{p}_{mc}\}_{  c \in    \mathcal{C}, m \in    \mathcal{M}_{\rm tr}}$ 
and $\{ \mathsf{h}_k \}_{k \in  \mathcal{K}_{\mathcal{B}}(b)}$.
The problem in~\eqref{eq:optimization_core_b_reformulation} can be easily transformed
to an equivalent  factor graph  (FG) and can be solved 
through the Max-Sum algorithm.

\begin{figure}[!t]
        \includegraphics[width=0.9\columnwidth]{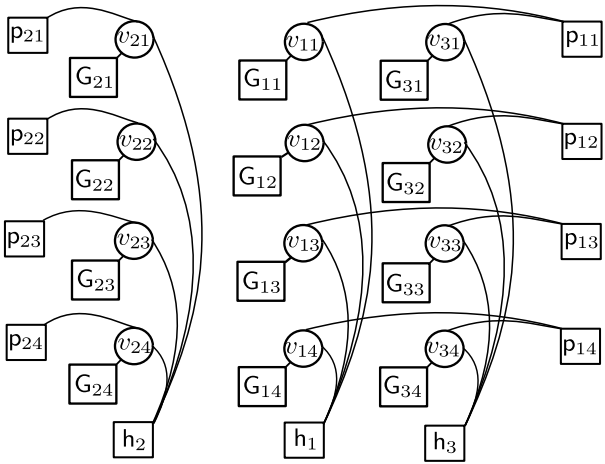}
\caption{A FG associated with the first core   (i.e., $b=1$) for the
  multi-cell BSN  of Fig.~\ref{fig:BackscatterSensorNetwork}. The FG assumes:
  (a) $C = 4$ frequency channels, (b) $M_{\rm tr}=2$ orthogonal training sequences of length 2,  and  (c)
 $\mathcal{K}_{11} = \{1,3\}$ and $\mathcal{K}_{12} = \{2\}$.}
\label{fig:FG_BackscatterSensorNetwork}
\end{figure}


A  FG  expresses factorizations   as   the one in Eq.~\eqref{eq:optimization_core_b_reformulation}, consisting of factor nodes
and variable nodes. Each factor node in the FG is connected  through an edge  to a variable node if the corresponding factor
has input the specific  variable. In the optimization problem~\eqref{eq:optimization_core_b_reformulation}
there exist $3$ types of factors:
\begin{itemize}
\item factors  $\{\mathsf{G}_{kc}: \forall (k,c)\in  \mathcal{K}_{\mathcal{B}}(b) \times \mathcal{C}\}$: each of them is connected to the corresponding  variable $v_{kc}$,
\item factors $\{ \mathsf{p}_{mc}\}_{ c \in    \mathcal{C}, m \in \mathcal{M}_{\rm tr}}$: each of them is connected to   variables $\{ v_{kc}\}_{k \in \mathcal{K}_{bm}}$, 
\item  factors  $\{ \mathsf{h}_k \}_{ k \in  \mathcal{K}_{\mathcal{B}}(b) } $: each of them is connected to   variables $\{ v_{kc}\}_{c \in \mathcal{C}}$.
\end{itemize}
Given the definition of the above factors and the assignment variable matrix $\mathbf{V}_{b}$,  a FG  can be constructed  for each core $b \in \mathcal{B}$.
 Each such FG corresponds to the Max-Sum factorization in~\eqref{eq:optimization_core_b_reformulation}.
An example of a FG associated with core $1$ at  the BSN of Fig.~\ref{fig:BackscatterSensorNetwork} is depicted in Fig.~\ref{fig:FG_BackscatterSensorNetwork}.

%


\begin{algorithm}[!b]
\caption{Max-Sum Algorithm}   \label{alg:max_sum_algo}
\textbf{Input:}   $\left\{ \overline{\mathtt{SINR}}_{kb}^{(c)}:  k\in \mathcal{K}_{\mathcal{B}}(b), c\in \mathcal{C}\right \}$
\begin{algorithmic} [1]
\State{ $n=0$, $\phi_{kc}^{(0)} = \rho_{kc}^{(0)}  = 0,~ \forall (k,c)\in \mathcal{K}_{\mathcal{B}}(b) \times \mathcal{C} ,$ $a\in [0,1)$ }
\While{termination criterion is {\bf not} reached}
\State{$n:=n+1$} 
\For{each $(k,c) \in \mathcal{K}_{\mathcal{B}}(b) \times \mathcal{C}$}
\State{$ \phi_{kc}^{(n)}  = \max \limits_{c' \in \mathcal{C} \backslash c}  \! \left\{ 
 -\rho_{kc'}^{(n-1)}  +  \mathsf{g}\!\left(\overline{\mathtt{SINR}}_{kb}^{(c')} \right) \right\} $}
 \State{$\phi_{kc}^{(n)}  := a\phi_{kc}^{(n-1)} + (1-a)\phi_{kc}^{(n)}$}
 \EndFor
 \For{each $m \in \mathcal{M}_{\rm tr}$ and each $(k,c) \in \mathcal{K}_{bm} \times \mathcal{C}$}
\State{   $\rho_{kc}^{(n)}  = \!   \left[   \max \limits_{k' \in \mathcal{K}_{bm} \backslash k} \! \left\{ 
-\phi_{k'c}^{(n)}  +  \mathsf{g}\!\left(\overline{\mathtt{SINR}}_{k'b}^{(c)} \right) \right\} \right]^+$}
 \State{$\rho_{kc}^{(n)} := a\rho_{kc}^{(n-1)} + (1-a) \rho_{kc}^{(n)}$}
\EndFor
\For{each $(k,c) \in \mathcal{K}_{\mathcal{B}}(b) \times \mathcal{C}$}
 \State{$ \chi_{kc}^{(n)} = \phi_{kc}^{(n)} +\rho_{kc}^{(n)}- \mathsf{g}\!\left(\overline{\mathtt{SINR}}_{kb}^{(c)}\right)$}
 \State{$\widehat{v}_{kc}^{(n)} = 1 $ if   $\chi_{kc}^{(n)} \leq 0 $, and  $0$, otherwise.}
\EndFor
\EndWhile
\end{algorithmic} 
{\textbf{Output:}  $\widehat{\mathbf{V}}_{b} =\left \{\widehat{v}_{kc}^{(n)}:   \forall (k,c) \in \mathcal{K}_{\mathcal{B}}(b) \times \mathcal{C}\right \}$}
\end{algorithm}

For a given FG,    the standard Max-Sum message-passing rules can be derived to find
the optimal   association matrix $\mathbf{V}_{b}$ that maximizes objective
function in~\eqref{eq:optimization_core_b} and satisfies the constraints~\eqref{eq:optimization_core_b_2}--\eqref{eq:optimization_core_b_5}.
\begin{theorem} \normalfont \label{theorem:update_rules}
 Algorithm~\ref{alg:max_sum_algo} applies    the  Max-Sum update rules 
associated with the   factorization  in Eq.~\eqref{eq:optimization_core_b_reformulation}.
%
\end{theorem}
 The proof of the above is provided in \cite[Appendix 5.6]{Alevizos_PhD_thesis:17}.
The overall procedure to solve the optimization problem in~\eqref{eq:optimization_core_b} is provided in   Algorithm~\ref{alg:max_sum_algo}.
The algorithm is executed for all cores $b \in \mathcal{B}$ in parallel.
It can be observed   that a damping technique with 
 an extra one-iteration-memory step is employed  at lines 6 and 10 of the algorithm.
 Damping technique is utilized  to prevent pathological oscillations \cite{Hes:04}.
Algorithm~\ref{alg:max_sum_algo}  terminates   either if a maximum number of iterations,  $n_{\rm max}$,  is reached,
or if the normalized max-absolute error (NMAE) between two consecutive soft-estimates, 
$
\frac{ \max_{(k,c) \in  \mathcal{K}_{\mathcal{B}}(b)  \times \mathcal{C} }  \left| \chi_{kc}^{(n)} -  \chi_{kc}^{(n-1)} \right| }
{\max_{(k,c) \in  \mathcal{K}_{\mathcal{B}}(b)  \times \mathcal{C} } \left| \chi_{kc}^{(n)} \right|} $,
is below a prescribed precision $\epsilon$. 
It is emphasized   that   Algorithm~\ref{alg:max_sum_algo} is fully parallelizable
with very simple update rules that require mainly  addition and comparison
operations.

 
Invoking the convergence results derived in  \cite{BaBoChZe:11} for general weighted matching problems, 
it follows that if the optimal  solution of  linear program (LP) associated with the relaxed version of 
problem~\eqref{eq:optimization_core_b} is integral (i.e., the  optimal solution belongs in $\mathds{B}^{|\mathcal{K}_{\mathcal{B}}(b)
| \times | \mathcal{C}|}$)  and unique, then Max-Sum algorithm converges to the exact solution after at most
$\mathcal{O}(C \, |\mathcal{K}_{\mathcal{B}}(b)|)$ iterations.

 Regarding per iteration computational cost of the proposed algorithm, it is not difficult to see  
that lines (4)--(7),  (8)--(11), and  (12)--(15)  require  $\mathcal{O}(C^2 |\mathcal{K}_{\mathcal{B}}(b)|)$, $\mathcal{O}(C\, |\mathcal{K}_{\mathcal{B}}(b)|^2)$,
and  $\mathcal{O}(C\, |\mathcal{K}_{\mathcal{B}}(b)|)$ arithmetic operations, respectively.
The algorithm iterates at most  $n_{\rm max}$ times, and thus, the overall worst-case computational complexity of Algorithm~\ref{alg:max_sum_algo}
becomes $\mathcal{O}(n_{\rm max} \,(C\, |\mathcal{K}_{\mathcal{B}}(b)|^2 + C^2 |\mathcal{K}_{\mathcal{B}}(b)|))$.


\section{Simulation Results}
\label{sec:sim_results}


A multi-cell BSN topology is considered, 
consisting of $B = 7$ cores and $K = 140$ tags. 
Cores  are placed in a cellular setting  \cite{Goldsmith:05} and the distance of neighboring
cores is $\sqrt{3} R_{\rm core}$, where $R_{\rm core} = 6$ meters.
The height of all cores 
 is $2$ meters. Tags  are scattered randomly in the
vicinity of cores. Their height is a uniform random variable in $[0,1]$.

The adopted  path-loss model is given by  \cite{Goldsmith:05}
$
\sigma_{bk}^2 = \sigma_{kb}^2 =    \left( {d_0}/{  \mathsf{d}_{bk}}\right)^{\nu_{bk}}  \left( {\lambda}/({4 \pi d_0})\right)^2,
\label{eq_large_scale_fading}
$
with   path-loss exponent  $\nu_{bk} =  2.1,~  \forall b \in \mathcal{B},\,\forall k \in \mathcal{K}_{\mathcal{B}}(b) $,
carrier wavelength  $\lambda \approx  0.3456$ m (UHF frequencies),
 and  $d_0=1$ m. Rician
 parameters are taken  $ \kappa_{kb}^{\rm u}= \kappa_{bk}^{\rm d} =10$ dB, $ \forall b \in \mathcal{B},\,\forall k \in \mathcal{K}_{\mathcal{B}}(b) $ 
 and noise variance  $\sigma_b^2 =     -174 +  {\rm NF} $ dBm,
 with noise figure  ${\rm NF} = 4$ dB.
For simplicity,  common backscatter reflection coefficients  are considered for all tags   with $\Gamma_{k,0} = 0.47$
and  $\Gamma_{k,1} = -0.54$, $k \in \mathcal{K}$.
The tag scattering efficiency is assumed common for all tags, given by  $ {\eta}_k = 0.2, ~\forall k \in \mathcal{K}$.
The cores' transmission power, $P_b$  is the same.
Each core is equipped with uniform linear arrays with $N_{\rm T}= 1$ transmit
and $N_{\rm R} = 4$ receive antennas assuming
known AoA and AoD relative to the neighboring tags. The number of available frequency sub-carrier in BSN is  $C=8$ 
 and $M_{\rm tr}=8$ orthogonal  training sequences are employed.
Each sub-channel is given by $\mathtt{f}^{(c)} =  \frac{2c}{T}   $, $c \in \mathcal{C}$
and  the symbol period is $T = 0.1$ msec. 
Finally, set $\mathcal{X}_{\rm tr} $ comprises of the columns of the
$M_{\rm tr}\times M_{\rm tr}$ Hadamard  matrix.


Using the   parameters  of previous paragraph, $J=10000$
SINR measurements are obtained  to estimate the average (long-term) SINR according to Eq.~\eqref{eq:avg_SINR_kb_c}.
The proposed algorithm~\ref{alg:max_sum_algo} is executed to obtain the optimal assignment of variables  $\{v_{kc}\}_{(k,c) \in \mathcal{K} \times \mathcal{C}}$,
 with parameters   $\mathsf{g}(x) = x, ~x\geq 0$,%
$n_{\rm max} = 100$, $\alpha = 0.05$, and $\epsilon = 10^{-5}$. The specific value for
the objective tries to maximize the sum of average SINRs, although other metrics could be 
  employed.

\begin{figure}[!t]
\centering
        \includegraphics[width=0.8\columnwidth]{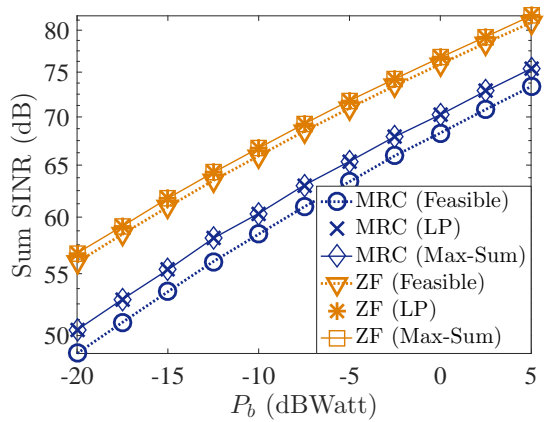}
\caption{Sum of SINRs for the Max-Sum algorithm, LP, and  orthogonal  channel allocation.}
\label{fig:sum_SINRs}
\end{figure}

Fig.~\ref{fig:sum_SINRs} compares the sum of average SINR performance across all tags
as a function of cores' transmission power.  As expected, for both ZF and MRC the proposed Max-Sum algorithm
offers the same optimal performance with the relaxed LP technique and both outperform the
channel allocation for which tags of the same cell use unique pairs $(m,c) \in \mathcal{M}_{\rm tr} \times \mathcal{C}$
(a feasible solution of problem \eqref{eq:optimization_core_b}). The performance of the latter
allocation is calculated across 1000 independent experiments.
It can be remarked that the gap between ZF and MRC is $5$-$10$ dB,
corroborating the intra-cell interference mitigation capabilities of ZF detector.

\begin{figure}[!t]
\centering
        \includegraphics[width=0.88\columnwidth]{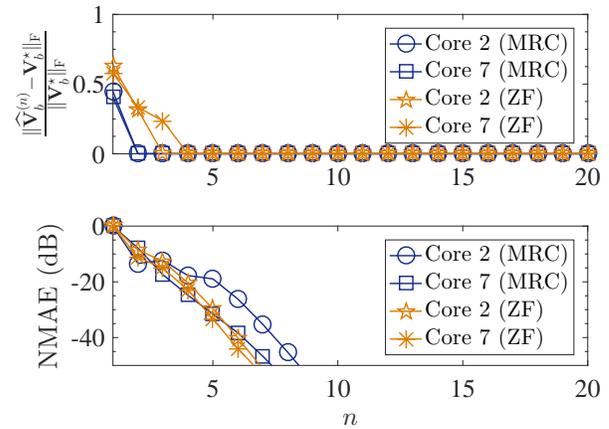}
\caption{Convergence rate and NMAE of soft-estimates for the Max-Sum algorithm executed at cores $2$ and $7$.}
\label{fig:outage_plots}
\end{figure}

  Fig.~\ref{fig:outage_plots} 
  shows how fast the proposed algorithm converges to the optimal $\mathbf{V}_{b}^{\star}$
and how many iterations are required  until the termination criterion is reached at cores $2$ and $7$.
For all cores the algorithm terminates after  $5$-$10$ iterations on average, and for all cases
the termination criterion of  soft-estimates NMAE   below $\epsilon$ was met.
Also, the algorithm converges to the optimal solution within $2$-$3$ iteration at all cores.
The above demonstrates the potential benefits of the proposed Max-Sum algorithm,
since per iteration complexity is   small and convergence is accomplished within   few steps.

%
%

Finally, one important question would be why one should use the proposed algorithm
instead of classic LP to solve the  studied optimization problem in~\eqref{eq:optimization_core_b}.
To this end, the proposed algorithm  was compared with   CVX
convex optimization solver \cite{cvx} in terms of average execution time across all cores.
For the simulations,       a    computer with 64-bit operating system
and Intel(R) Core(TM) i7-3540 and CPU at 3 GHz was used.
The proposed algorithm was implemented 
with a custom MATLAB script,
while the solution of LP relaxed problem was obtained by CVX solver.
The average execution time, averaged across several transmit powers, for the proposed
Max-Sun algorithm was $0.03$ sec and $0.033$ sec for ZF and MRC detectors, respectively,
whereas, for the LP program with CVX solver 
required $0.424$ sec and $0.5$  for ZF and MRC detectors, respectively.
This shows at least an order of magnitude improvement.

%

\section*{Acknowledgment}
This research is implemented through the Operational Program ``Human Resources Development, Education and Lifelong Learning'' and is co-financed by the European Union (European Social Fund) and Greek national funds.

%


\end{document}